\documentclass[letterpaper,twocolumn,pra,
aps,showpacs,superscriptaddress,floatfix]{revtex4-2}

\usepackage[latin1]{inputenc}
\usepackage{bbm}
\usepackage{bm}
\usepackage[usenames]{color}
\usepackage{multirow}
\usepackage{amssymb}
\usepackage{amsbsy}
\usepackage{mathtools}
\usepackage{amsmath}
\usepackage{stmaryrd}
\usepackage{graphicx}
\usepackage{epsfig}
\usepackage{placeins}
\usepackage{bbold}
\usepackage{braket}
\usepackage{blindtext}
\usepackage{subcaption}
\usepackage[colorlinks,linkcolor=blue,citecolor=blue,urlcolor=blue]{hyperref}

\usepackage{scalerel}

\usepackage{tikz}
\usetikzlibrary{calc}
\usetikzlibrary{patterns}
\usetikzlibrary{svg.path}
\definecolor{orcidlogocol}{HTML}{A6CE39}
\tikzset{
  orcidlogo/.pic={
    \fill[orcidlogocol] svg{M256,128c0,70.7-57.3,128-128,128C57.3,256,0,198.7,0,128C0,57.3,57.3,0,128,0C198.7,0,256,57.3,256,128z};
    \fill[white] svg{M86.3,186.2H70.9V79.1h15.4v48.4V186.2z}
                 svg{M108.9,79.1h41.6c39.6,0,57,28.3,57,53.6c0,27.5-21.5,53.6-56.8,53.6h-41.8V79.1z M124.3,172.4h24.5c34.9,0,42.9-26.5,42.9-39.7c0-21.5-13.7-39.7-43.7-39.7h-23.7V172.4z}
                 svg{M88.7,56.8c0,5.5-4.5,10.1-10.1,10.1c-5.6,0-10.1-4.6-10.1-10.1c0-5.6,4.5-10.1,10.1-10.1C84.2,46.7,88.7,51.3,88.7,56.8z};
  }
}

\newcommand\orcid[1]{\!%
  \href{https://orcid.org/#1}{%
    \mbox{%
      \scaleto{%
        \begin{tikzpicture}[yscale=-1,transform shape]
          \pic{orcidlogo};
        \end{tikzpicture}
      }{8pt}%
    }%
  }%
}

\makeatletter

\begin{document}
\title{Examination of classical simulations for Heisenberg-Langevin equations for spin-1/2}
\author{Scott D. Linz~\orcid{0009-0005-7777-7955}}
\affiliation{Department of Mathematics/Computer Science/Physics, University of Osnabr\"uck, D-49076 
Osnabr\"uck, Germany}
\email{sclinz@uni-osnabrueck.de}
\author{Jochen Gemmer~\orcid{0000-0002-4264-8548}}
\affiliation{Department of Mathematics/Computer Science/Physics, University of Osnabr\"uck, D-49076 
Osnabr\"uck, Germany}

\date{\today}

\begin{abstract}
A system of spins coupled to a bath is a traditional setup in open quantum systems. Through Heisenberg's equation, the spin dynamics can be modeled by a set of first-order differential equations. Interpreting the terms as colored noise and non-Markovian damping, one can write them as quantum-mechanical Heisenberg-Langevin (HL) equations. These are notoriously difficult to solve because of the high dimensionality of the Hilbert space. Classical generalized Langevin equations, involving non-Markovian damping and colored noise, are well understood and can be treated numerically with relative ease. Thus, a classical ansatz can be made by substituting quantum expectation values with classical functions. This allows the application of standard methods developed for classical stochastic dynamical systems to tackle spin dynamics. However, this approach is uncontrolled and should be benchmarked against known quantum dynamics. In this investigation, a Hamiltonian for spin dynamics is modified to obtain a setup analogous to the Weisskopf-Wigner (WW) theory of spontaneous emission, enabling a comparison of the results. This will be compared for $T=0$ and with a slight adaptation in the high-temperature limit.
\end{abstract}

\maketitle

\section{Introduction}

In the early 20th century, physicists' understanding of physical phenomena underwent profound transformations, particularly in regimes far away from the classical domain. Einstein's theory of relativity revolutionized our understanding of the high-energy regime, while quantum mechanics provided a new description of the microscopic world. Since standard classical physics describes the everyday world familiar to everyday experience, Newtonian mechanics should emerge as a limiting case of each of these theories.\\
In special relativity, this transition is simple. One simply takes the limit $c \rightarrow \infty$. Thereby, the Lorentz group reduces to the Galilei group, recovering Newtonian mechanics. In quantum mechanics, however, the classical limit is much more delicate. Setting $\hbar = 0$ does not, in general, yield classical dynamics. Ehrenfest's theorem laid the groundwork for understanding this transition, but it features a subtle limitation. The dynamics of the quantum expectation values generally do not coincide with the classical trajectories, since $\langle V(x) \rangle \neq V(\langle x \rangle)$. Unlike position and momentum, spins with low quantum numbers lack a direct classical analog, as they are intrinsic angular momenta that do not arise from classical trajectories, making attempts to describe them classically particularly difficult and often heuristic. Various classical approximations have been developed to bridge this gap, each with varying degrees of validity depending on the system and the limit of interest \cite{gaioli1998classical, morales2021semi,PhysRevB.104.054415,PhysRevLett.110.147201,schachenmayer2015many,PhysRevB.109.014427}. Notably, quantum effects become increasingly prominent at low temperatures, a limit that is of central importance in experimental condensed matter physics.\\
A useful tool for describing open quantum systems are Heisenberg-Langevin (HL) equations, which allow rewriting the quantum dynamics of operators as a stochastic differential equation \cite{scully1997quantum}. In this formulation, the influence of an external environment appears as a combination of damping and colored noise, which are linked through a fluctuation-dissipation relation. For bosonic systems, the transition from quantum to classical Langevin dynamics is relatively well understood \cite{cortes1985generalized}. For spin systems, a classical calculation can be made by replacing quantum operators with classical vector components. This approximation lacks a small parameter and therefore does not systematically approach the quantum dynamics in any known limit, making it uncontrolled.\\
This work aims to investigate and examine the validity of such a classical calculation of spin dynamics. Anders et al. \cite{anders2022quantum} proposed a HL equation for spin systems coupled to an external bath with related works being \cite{hartmann2023anisotropic, hogg2024enhanced, hogg2024tutorial, cerisola2024quantum}. To calculate the time evolution for expectation values of a spin system, a classical ansatz was made. We aim to investigate the validity of this approach by benchmarking it against the dynamics of a system with well-understood quantum dynamics. Here, a simplified version of the HL, involving only a single spin-1/2, introduced in Section \ref{Sec2}, will be considered. Its dynamics for the zero temperature limit will be compared to those of the well-established Weisskopf-Wigner (WW) theory of spontaneous emission, which describes a single spin coupled to a bath at zero temperature. In Section \ref{Sec3}, we demonstrate how the Hamiltonians from the work by Anders et al. \cite{anders2022quantum} and the WW setup can be adapted to be analogous by tuning parameters. In section \ref{Sec4} we compare the results of a thermal average of the classical calculation of the HL to the known quantum dynamics at $T=0$. Here, the coupling will be varied to approach the scenario of approximately Markovian damping and also a setup far from Markovianity. Furthermore, in Section \ref{Sec5}, the high-temperature quantum spin dynamics can be computed, although only for the strictly Markovian case. This result will be compared to the high-temperature dynamics from the classical HL examined in this paper in order to examine how the dynamics compares when the bath is occupied. 

\section{Heisenberg-Langevin equation for spin dynamics}\label{Sec2}

To simulate spin precession in a material, a generalized Landau-Lifshitz-Gilbert equation in the form of a HL equation has been proposed by Anders et al. \cite{anders2022quantum}, being derived from microscopic principles. This HL equation in its general form features numerous spins of an arbitrary spin quantum number in a magnetic field interacting with each other. Every spin is additionally coupled to its own bath. For the purposes of this work, we restrict our attention to a single spin-1/2 with its bath and thereby neglect interactions between different spin sites.\\
The derivation begins with the standard setup for open quantum systems by defining a Hamiltonian being composed of three parts:
\begin{equation} \label{Ham}
    \hat{H} = \hat{H}_S + \hat{H}_B + \hat{V}_\text{Int},
\end{equation}
where $\hat{H}_S$ describes the system (a single spin), $\hat{H}_B$ is modeled as a bath of harmonic oscillators, and $\hat{V}_\text{Int}$ captures the interaction between the spin and the bath.
The system Hamiltonian is given by
\begin{equation}\label{System}
    \hat{H}_S = - \hat{\mathbf{S}} \cdot \mathbf{B}_\text{ext},
\end{equation}
being just the Zeeman interaction of a spin-1/2 with an external magnetic field $\mathbf{B}_\text{ext}$. Here, the gyromagnetic ratio $\gamma=1$ and thereby $\hbar=1$ are set to unity, allowing for a unit-free calculation. The bath is modeled as a continuum of independent harmonic oscillators:
\begin{equation}\label{Bath}
    \hat{H}_B = \frac{1}{2} \int_0^\infty d \omega \left[ \hat{\mathbf{X}}_\omega^2 + \hat{\mathbf{P}}_\omega^2 \right],
\end{equation}
where $\hat{\mathbf{X}}_\omega$ and $\hat{\mathbf{P}}_\omega$ are the position and momentum operators for each oscillator mode of frequency $\omega$.\\
The spin-bath interaction is mediated via a frequency-dependent coupling tensor $\mathcal{C}_\omega$, and takes the form of
\begin{equation}\label{Int}
    \hat{V}_\text{Int} = - \hat{\mathbf{S}} \cdot \int_0^\infty d \omega \mathcal{C}_\omega \hat{\mathbf{X}}_\omega.
\end{equation}
where both the system's spin operator and the position operator of the continuous bath modes enter linearly. For this investigation, it suffices to consider a tensor $\mathcal{C}_w$ composed of two factors
\begin{equation}
    \mathcal{C}_\omega = c_\omega \mathcal{A},
\end{equation}
where the function $c_\omega$ governs the spectral density of the coupling for any frequency $\omega$ and $\mathcal{A}$ governs the anisotropy and strength of the coupling.\\
Anders et al. \cite{anders2022quantum} arrive at first-order differential equations for the spin in the form of an exact HL equation containing stochastic noise and dissipative non-Markovian damping:
\begin{equation} \label{LH}
    \frac{d \hat{\mathbf{S}}}{dt} = \frac{1}{2} \hat{\mathbf{S}} \times \left[ \mathbf{B}_\text{ext} + \hat{\mathbf{b}}(t) + \int_0^t dt' \mathcal{K}(t - t') \hat{\mathbf{S}}(t') \right] + \text{h.c.},
\end{equation}
where h.c. denotes the Hermitian conjugate of the entire right-hand side to ensure that the equation preserves operator Hermiticity. This expression contains three terms:\\
$\bullet$ the unperturbed Zeeman interaction with the spin, where the direction of $\mathbf{B}_\text{ext}$ determines the ground state.\\
$\bullet$ a stochastic magnetic field $\hat{\mathbf{b}}(t)$ representing colored quantum noise,\\
$\bullet$ a damping term governed by a memory kernel $\mathcal{K}(t - t')$ that encodes (in general non-Markovian) dissipation.\\
Importantly, the HL equation as given by Eq. (\ref{LH}) preserves the norm of the spin vector, ensuring that the dynamics of a spin initially lying on the Bloch sphere remains confined to its surface. It should be noted that this feature is preserved when switching from a quantum mechanical to a classical description of the spin. Therefore, resulting classical trajectories are guaranteed to be within a physically justified interval.\\
The initial state is assumed to be the product state of the system $\varrho_S(0)$ and the bath $\varrho_B(0)$, which means that the density matrix takes the form $\varrho_\text{Tot}(0) = \varrho_S(0) \otimes \varrho_B(0)$. The noise term $\hat{\mathbf{b}}(t)$ in (\ref{LH}) arises from the initial conditions of the bath oscillators and is given by
\begin{equation}
    \hat{\mathbf{b}}(t) = \int_0^\infty d \omega \frac{1}{\sqrt{2 \omega}} \mathcal{C}_\omega \left[ a_\omega(0) e^{-i \omega t} + \text{h.c.} \right],
\end{equation}
where $a_\omega(0)$ are the initial annihilation operators of the bath modes. 
This noise operator has zero mean as the bath is assumed to be in thermal equilibrium initially, but non-zero time correlations are present, as will be discussed below.\\
The memory kernel in (\ref{LH}) can be written out as
\begin{equation}\label{Kernel}
    \mathcal{K}(t - t') = \int_0^\infty d \omega \frac{\mathcal{C}_\omega \mathcal{C}_\omega^T}{\omega} \sin[\omega(t - t')],
\end{equation}
quantifying how the history of the system's states affects the current state of the system-bath interaction.\\
Within the context of HL equations, the coupling between the system and bath adds stochastic fluctuations into the dynamics of the system. These fluctuations are characterized not only by their amplitudes, but also by their temporal correlations. 
The spectral profile of the system-bath interaction strongly affects the dynamics. Lorentzian coupling, suggested by Anders et al. \cite{anders2022quantum}, models a peaked spectral density centered at $\omega_0$ with width $\Gamma$. The square of such a spectral density is given by:
\begin{equation}\label{Coupling}
    c_\omega^2 = \frac{2 \Gamma}{\pi} \frac{\omega^2}{(\omega_0^2 - \omega^2)^2 + \omega^2 \Gamma^2}.
\end{equation}
This spectral density shall be used for the remainder of this paper. Calculating the memory kernel tensor in the time domain yields
\begin{equation}
    \mathcal{K}(\tau) = \Theta(\tau) \mathcal{A} \mathcal{A}^T \exp \left(-\frac{\Gamma \tau}{2} \right) \sin(\Omega \tau),
\end{equation}
where $\Omega=\sqrt{\omega_0^2-\Gamma^2/4}$ and $\Theta(\tau)$ is the Heavyside step function. This reflects a damped oscillatory memory, characterized by a frequency $\Omega$ and a damping rate $\Gamma / 2$.\\
Simulating the full quantum dynamics of Eq. (\ref{LH}) is computationally expensive, as the Hilbert space grows exponentially with system size. An ansatz proposed is to replace the initial expectation values of the spin operators with vector components and let them evolve classically. In this classical ansatz, the Hermitian conjugate terms in Eq.(\ref{LH}) become redundant and the prefactor of $1/2$ is thereby removed, yielding a system of differential equations of the form
\begin{equation}\label{classHL}
    \frac{d \mathbf{S}}{dt} = \mathbf{S} \times \left[ \mathbf{B}_\text{ext} + \mathbf{b}(t) + \int_0^t dt' \mathcal{K}(t - t') \mathbf{S}(t') \right],
\end{equation}
where it should be noted that calculating the equations of motion for a classical spin using Poisson brackets yields the same first-order differential equations.
What makes this approach efficient is that it is computationally inexpensive and can be treated with the standard tools for solving classical Langevin equations. Furthermore, the property of the dynamics being confined to the Bloch sphere also holds for the classical calculation and guarantees physical plausibility of the resulting trajectories.\\
The noise is generated via the following instruction. First, the vector components of Gaussian random numbers are chosen, called $\xi_i$, with mean $\langle \xi_i(t)\rangle=0$ and variance $\sigma^2_{\xi_i}(t)=\frac{1}{\Delta t}$, to ensure that the noise accurately approximates continuous white noise in the limit $\Delta t \rightarrow 0$. These are generated for each spatial dimension at the times examined. Consequently, the method for generating colored noise can thereby be succinctly expressed by making use of the Fourier transform $\mathcal{F}$ and the inverse Fourier transform $\mathcal{F}^{-1}$ as
\begin{align}\label{noise1}
\mathbf{b}(t) &= \mathcal{A} \mathcal{F}^{-1}\left(\sqrt{P(\omega)} \tilde{\boldsymbol{\xi}}(\omega)\right)\\\label{noise2}
\tilde{\boldsymbol{\xi}}(\omega) &= \mathcal{F}\{ \boldsymbol{\xi}(t)\},
\end{align}
where $P(\omega)$ is the scalar power spectrum. In this examination, we will focus on a quantum mechanical power spectrum that takes the form
\begin{equation}\label{Power}
    \tilde{P}(\omega) = \frac{\pi c_\omega^2}{2\omega} \coth \left( \frac{\omega}{2T}\right)
\end{equation}
where the prefactor results from quantizing the bath oscillators, and the term $\coth(\dots)$ interpolates between the zero-point quantum case for temperature $T \rightarrow 0$, where $\coth(\dots) \approx 1$ and the high-temperature classical limit $\coth(\dots) \approx \frac{T}{\omega}$ for $T \gg \omega$. This smoothly captures whether quantum and thermal fluctuations dominate at different energy scales.\\
After fixing the procedure to generate colored noise, the setup for the classical description of the HL is complete. Precisely this classical treatment of spin dynamics is what the study at hand aims to scrutinize. In order to achieve this in \ref{Sec3}, \ref{Sec4} and \ref{Sec5} we will compare the trajectories of the classical calculation described above to the known quantum dynamics of a spin-1/2 coupled to a bath of harmonic oscillators.

\section{Analogy to generalized Weisskopf-Wigner setup}\label{Sec3}

After a description of the procedure proposed by Anders et al. \cite{anders2022quantum} to simulate spin dynamics, we aim to assess the validity of the classical HL equation. To do this, we will benchmark it against a setup with an easily solvable quantum dynamics. A prime candidate for such a comparison is the Weisskopf-Wigner (WW) model for spontaneous emission in a two-level system, as presented in the textbook by Scully and Zubairy \cite{scully1997quantum}.\\
Within the framework of the WW model, the two-level system is modeled as a spin-1/2. To match the HL model to the WW setup, we must adjust both the system Hamiltonian and the system-bath interaction terms accordingly. The modified Hamiltonian and its components shall be denoted by the superscript $M$.\\
To bring the Hamiltonian from which the HL equation arises into alignment with the WW framework, the external magnetic field in (\ref{System}) must have only a $z$-component. To make the spin point in the negative $z$-direction the ground state, the external magnetic field must be antiparallel to the excited spin, i.e. $\mathbf{B}_\text{ext} = -B_\text{ext} \mathbf{e}_z$. This simplifies the system Hamiltonian to
\begin{equation}\label{ModSys}
    \hat{H}^M_S = B_\text{ext} \hat{S}_z,
\end{equation}
To mimic the WW interaction, the resonance frequency must be chosen to have the same magnitude as the external magnetic field $\omega_0 = -B_\text{ext}$ in Eq. (\ref{Coupling}). Thus, the Zeeman splitting matches the system's transition frequency, which is precisely the system Hamiltonian used in WW theory.\\
In both cases, the bath is modeled as a continuous collection of harmonic oscillators and requires no further modification.\\
To encode a simple structure parallel to the WW setup, we choose an anisotropic coupling tensor for the HL Eq. (\ref{LH}) such that in Eq. (\ref{Int}) only the $x$-component of the spin couples to the $x$-component of the bath:
\begin{equation}
    \mathcal{C}_\omega = c_\omega
    \begin{pmatrix}
        \sqrt{\alpha} & 0 & 0 \\
        0 & 0 & 0 \\
        0 & 0 & 0
    \end{pmatrix}.
\end{equation}
A key approximation in WW theory is the rotating wave approximation (RWA), which neglects terms in the interaction Hamiltonian that do not conserve the excitation number. The RWA becomes increasingly accurate in the limit of large energy splitting, or in this particular case, large magnetic field strength.\\
Therefore, we rewrite the interaction term, given by Eq. (\ref{Int}), using spin raising and lowering operators and bosonic creation/annihilation operators. Thus, we split $\hat{S}_x = \hat{S}_+ + \hat{S}_-$ and $\hat{X}_\omega = \hat{a}_\omega^\dagger + \hat{a}_\omega$. The interaction can then be written as:
\begin{align}\label{ModInt}
\hat{V}^M_\text{Int} \approx - \frac{1}{2\sqrt{2}} \left( \hat{\sigma}_+ \int_0^\infty d\omega \frac{c_\omega}{\sqrt{\omega}} \hat{a}_\omega \hat{\sigma}_- \int_0^\infty d\omega \frac{c_\omega}{\sqrt{\omega}} \hat{a}_\omega^\dagger \right).
\end{align}
This form preserves the excitation number and thus matches the interaction structure of WW theory.  Thus, the free parameters of the Hamiltonian, as given in Eqs. (\ref{Ham})-(\ref{Int}), are adjusted to yield a scenario that approximates WW theory.\\
Now, the results of a fully quantum mechanical treatment of the WW setup for generalized coupling spectra will be briefly laid out. The Hamiltonian for WW consists of a sum containing the modified system and interaction, as given by Eqs. (\ref{ModSys}) and (\ref{ModInt}) as well as the unmodified bath given by (\ref{Bath}):
\begin{equation}
    \hat{H}^M = \hat{H}_S^M + \hat{H}_B + \hat{V}_\text{Int}^M.
\end{equation}
Within the derivation of the dynamics of this spin-boson model, the coupling between the spin and the bosonic bath can be freely chosen to suit the physical scenario. The spectral distribution of the coupling is chosen to be Lorentzian along the lines of Eq. (\ref{Coupling}). Specifically, applying a Lorentzian spectral coupling leads to a Volterra integro-differential equation for the excitation amplitude $\varphi(t)$ of the spin
\begin{align}\label{kernel1}
\dot{\varphi}(t) &= - \int_0^t dt'K(t-t')\varphi(t')
\end{align}
involving a memory kernel $K(\tau)$ in the form of
\begin{align}
\label{kernel2}
K(\tau) &=  \frac{\alpha}{8}\int_0^\infty d\omega \frac{c_\omega^2}{\omega} e^{i(\omega_0-\omega)\tau}.
\end{align}
For the investigation in this paper, the system shall always start in the fully excited state ($\hat{S}_z(0) = +1$). The bath will initially be in the vacuum state, which corresponds to the limit $T=0$. The expected quantum dynamics can be expressed by the excitation amplitude $\varphi(t)$ calculated with Lorentzian coupling according to Eq. (\ref{Coupling}) and is directly connected to the quantum spin dynamics from WW theory via rescaling
\begin{equation}\label{ca}
    \langle \hat{S}_z(t) \rangle = 2|\varphi(t)|^2-1.
\end{equation}
Here, the amplitude maps to the spin expectation value along the quantization axis.\\
In standard WW theory, the coupling is assumed to be frequency-independent over the narrow spectral region where the bath density of states and coupling strength are significant. This yields a Markovian time evolution, where the memory kernel $K(\tau)$ becomes a delta function, and the decay is purely exponential.\\
To identify limiting cases in which the classical method can be compared with analytically predictable dynamics, one must consider characteristic decay time scales. This will lead to the application of the notion of Markovianity of the dynamics, in which memory decays rapidly. In contrast, we will also consider strongly non-Markovian dynamics, in which the memory kernel decays on a time scale comparable to or longer than the characteristic decay time of the excitation amplitude. In this regime, the system's past continues to influence the present evolution, invalidating the Markov approximation and leading to pronounced memory effects in the dynamics. Hence, we will analyze the characteristic time scales for the decay of both the memory and the dynamics of the system. Around the resonance, the coupling function Eq. (\ref{Coupling}) behaves as a Lorentzian with a width of $\Gamma/2$ with small corrections of order $\omega^{-1}$. A Lorentzian's Fourier transform in time yields an exponential decay $K(\tau) \sim e^{-i\omega_0t}e^{-\Gamma/2 t}$. Thus, a characteristic decay time for the memory is given by $\tau_K = 2/\Gamma$.
Next, we introduce the characteristic decay time scale $\tau_\varphi$ for the excitation amplitude, defined as the inverse of the effective decay rate associated with the resonance. This rate is given by the maximum of the Lorentzian. Introducing a short-hand notation that will also be relevant below by defining
\begin{equation}\label{decayrate}
\lambda=\frac{\alpha}{2\Gamma \omega_0},
\end{equation}
we receive the characteristic time scale for the decay of the excitation amplitude as
$\tau_\varphi=2/\lambda$. 
Thus, we may define a Markovianity parameter $\mu_0=\tau_\varphi/\tau_K$ at $T=0$ that relates how the time scales differ and thereby measures how deep the dynamics are in the Markovian regime. Taking the characteristic time scales for the memory and the dynamics of the system, we can define two limiting cases for large and small $\mu_0$, respectively
\begin{align}
\mu_0=\frac{2 \Gamma^2 \omega_0}{\alpha} & \qquad \text{with} \\
\mu_0\gg 1 & \qquad \text{Markovian regime} \label{Mcond} \\
\mu_0 \sim 1 & \qquad \text{strongly non-Markovian.} \label{AMcond}
\end{align}
This allows us to tune parameters to match the aforementioned scenarios, whose strict limiting cases will be discussed below. For a strictly Markovian setup, which is the standard WW decay with $\mu_0 \rightarrow\infty$, the $z$-component of the spin undergoes an exponential decay
\begin{equation}\label{WWres}
    \langle \hat{S}_z(t) \rangle = 2e^{-\lambda t} - 1.
\end{equation}
with the decay constant $\lambda$ given by Eq. (\ref{decayrate}).
For $\mu_0 \sim 1$, we obtain the highly non-Markovian case. This parameter choice approximates the opposite of the Markovian scenario. This means that one approaches the case where the Loretzian spectral density approximates a delta peak at $\omega_0$, if one lets $\Gamma \rightarrow 0$ and $\Gamma/\alpha$ be constant. Here, oscillatory dynamics of the $\langle \hat{S}_z (t) \rangle$ with a damping controlled by $\Gamma$ is expected. Here $\Gamma \rightarrow 0$ with $\Gamma/\alpha = $const. would imply no damping. Thus, the overall behavior of the reference curves with which the classically computed HL equation will be compared is established.

\section{Comparison at T=0}\label{Sec4}

\begin{figure}[t]    
    \includegraphics[scale=0.7]{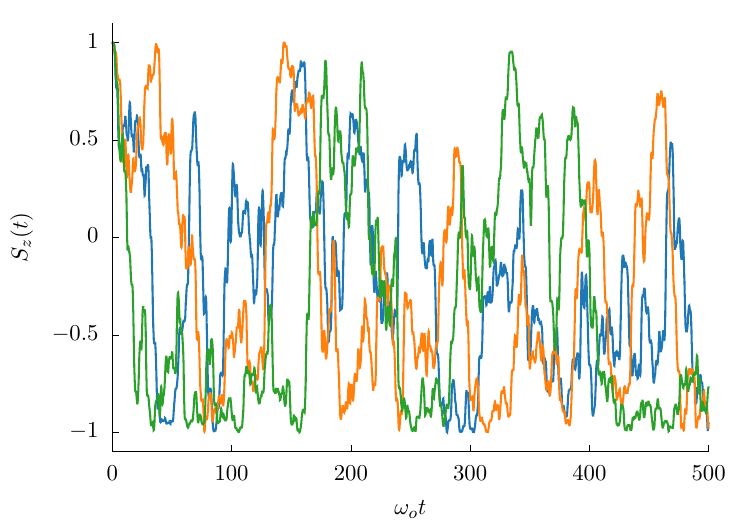}
    \caption{Three stochastic realizations of $S_z(t)$ under quantum zero-point noise. Each trajectory begins at $S_z(0) = +1$ and evolves stochastically. This figure demonstrates the need to average over many stochastic realizations to obtain a smooth curve, where the decay rate and long-time average can be analyzed. Here we chose the parameter set as $\Gamma=7.5$, $\omega_0=5$ and $\Gamma/\alpha=1$.}
    \label{fig:trajectories}
\end{figure}

This section aims to simulate the HL given by (\ref{classHL}) at $T=0$ and compare it to the quantum mechanical dynamics, both in the Markovian case and far away from the Markovian case.\\
The HL equation will be calculated classically with quantum zero-point noise arising from Eq. (\ref{Power}) and implemented along the lines of Eqs. (\ref{noise1}) and (\ref{noise2}).
\begin{figure}[t] 
    \includegraphics[scale=0.7]{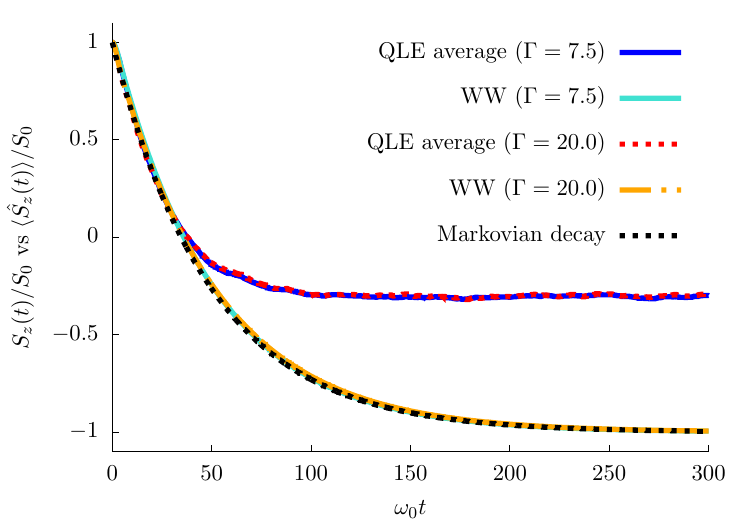}
    \caption{This figure shows the ensemble-averaged decay of $S_z(t)$ under quantum noise averaged over $5,000$ realizations as well as the WW result with Lorentzian coupling. The parameters were chosen as $\Gamma/\alpha=1$ and $\omega_0=5$ in both cases, with two distinct choices for the remaining free parameter. These are $\Gamma_1=7.5$ and $\Gamma_2=20$. The solid lines result from the configuration $\Gamma_1$, whereas the dashed red and orange curves arise from the configuration $\Gamma_2$. Both parameter choices produce Markovian dynamics, where the second configuration is deeper in the Markovian limit with $\mu_{0,2}>\mu_{0,1}$. Comparing the dashed black line, which is the exponential decay resulting from the strictly Markovian WW calculation according to Eqs. (\ref{WWres}) and (\ref{decayrate}), one can see that both configurations excellently approximate this limit. 
    }
    \label{fig:average_decay}
\end{figure}
\begin{figure}[t]
    \includegraphics[scale=0.7]{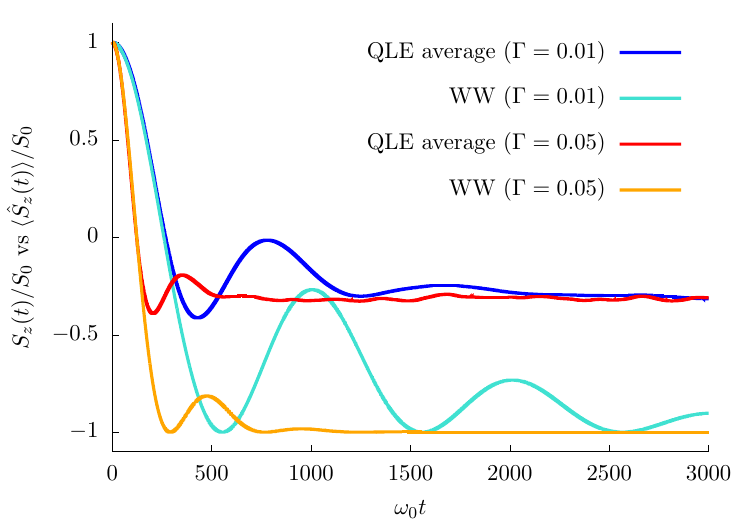}
    \caption{Here, the ensemble-averaged decay of $S_z(t)$ is plotted over an average of $5,000$ realizations alongside the generalized WW result with Lorentzian coupling. The parameters were chosen so that $\Gamma/\alpha=1$, $\omega_0=5$ in both cases, and the remaining free parameter was varied with $\Gamma_3=0.01$, blue and turquoise, and $\Gamma_4=0.05$, red and orange. Damped oscillatory dynamics becomes clearly visible in both cases. The fourth configuration is less damped since $\mu_{0,3}<\mu_{0,4}$, where $\alpha/\Gamma=$ const..
    }
    \label{fig:average_decay_2}
\end{figure} 
Quantum WW data were generated by computing the Volterra equation (\ref{kernel1}) for both the Markovian limit and the limit far away from Markovian dynamics. In both parameter regimes, a coupling of (\ref{Coupling}) is chosen to guarantee comparability with the classical HL simulation. The dynamics is scaled according to (\ref{ca}) to show the evolution of the expectation value of $\langle S_z(t) \rangle$. The Volterra equation was solved by numerically computing the integral for the memory kernel Eq. (\ref{kernel2}) in each time step, up to a point where the additional contributions become negligible. Then, using the Euler forward method, the following time step is generated from the previous one.\\
To numerically evaluate the classical dynamics with quantum noise, we simulate an ensemble of trajectories where the spin evolves under the influence of colored noise generated from the quantum power spectrum in Eq. (\ref{Power}). To generate time-evolution data under this zero-point noise, we used the open-source package \texttt{SpiDy} \cite{scali2024spidy}, implementing noise generation according to Eqs. (\ref{noise1}) and (\ref{noise2}). In this section, the resonance frequency is set to $\omega_0=5$ and in accordance with WW theory as discussed above $B_\text{ext}=-5$ so that spin down is the ground state. 
Furthermore, $\Gamma/\alpha=1$ will be chosen for the rest of this section. This removes one free parameter. The parameter $\Gamma$ will be varied with $\alpha$ taking corresponding values. Figure \ref{fig:trajectories} shows three example trajectories of $S_z(t)$, where the effect of quantum noise is clearly visible: the trajectories fluctuate and decay away from the unstable fixed point. However, since noise dominates the individual classical trajectories, the ensemble averages over many realizations, as suggested by Anders et al. \cite{anders2022quantum}, will be compared to the expectation value gained from the WW prediction.\\
The following figures, Fig. \ref{fig:average_decay}, for the Markovian limit and Fig. \ref{fig:average_decay_2}, for the limit far away from Markovianity, compare the time evolution predicted by the averaged classical HL equation with quantum noise to the decay expected from the WW model.\\ 
As shown in Fig. \ref{fig:average_decay}, according to Eq. (\ref{Mcond}), we first set $\Gamma_1=7.5$, producing a Markovianity parameter $\mu_{0,1}=75$, and in the second case we set $\Gamma_2=20$ yiedling $\mu_{0,2}=200$. Thus, the second parameter set is even closer to the Markovian limit. The averaged trajectory $\langle S_z(t) \rangle$ decays approximately exponentially, but does not reach the ground state $S_z = -1$. Instead, it saturates in the steady state $\bar{S}_z(\infty) \approx -0.31$, due to persistent fluctuations. This is precisely the steady state for the thermal average as predicted in \cite{anders2022quantum}. Note that this discrepancy is expected to become less pronounced if the calculation involves larger spins. However, the rate of decay initially matches the prediction from the strictly Markovian WW set-up and the trajectories computed with Lorentzian coupling in the Markovian regime almost perfectly, until it plateaus around the steady state. Computing the dynamics with a classical power spectrum and thus without any noise at $T=0$ leads to a dynamics being frozen at the initial conditions, as demonstrated in the appendix.\\
Secondly, we also considered two parameter configurations that are not in the Markovian limit according to Eq. (\ref{AMcond}). Here, the solutions to (\ref{kernel1}) become damped oscillations.
The third configuration with $\Gamma_3=0.01$ yields $\mu_{0,3}=0.1$, which is the furthest away from the Markovian limit of all cases examined in this paper. Here, the damped oscillations are clearly visible. The fourth set, where $\Gamma_4=0.05$ yields $\mu_{0,4}=0.5$, also features notable oscillations, although the damping is more pronounced. As can be seen in Fig. \ref{fig:average_decay_2}, the comparison still features significant differences. The steady state to which the dynamics converges still differs, being $\langle \hat{S}_z(\infty) \rangle = -1$ for the quantum calculation and again $\bar{S}_z(\infty) = -0.31$ for the classical simulation. The rate of decay is almost identical initially; however, the frequency of the oscillations is slightly larger for classical HL as compared to the non-Markovian WW time evolution.

\section{High-Temperature Comparison}\label{Sec5}

\begin{figure*}
\centering
\begin{subfigure}{5.5cm}
\centering
\includegraphics[width=\textwidth]{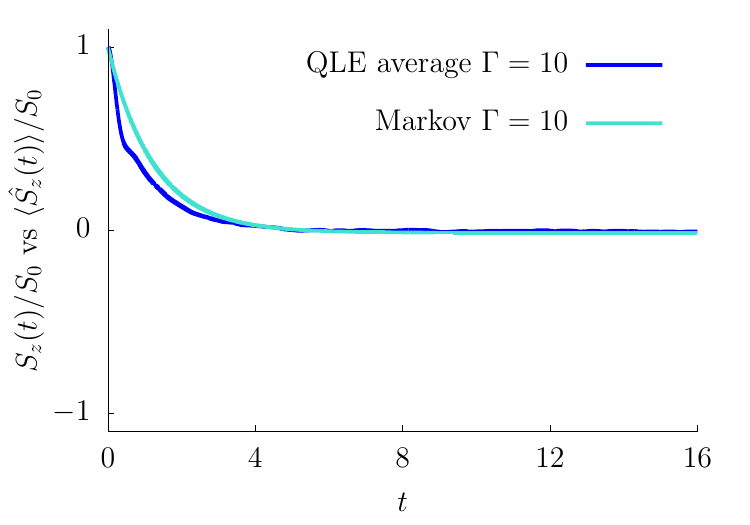}
\caption{Here, the parameters for the classical HL simulation are $\Gamma_1=10$, $\omega_0=5$ and $\alpha=1$ averaged over $25,000$ trajectories.}
\label{fig:sub1}
\end{subfigure}
\hfill
\begin{subfigure}{5.5cm}
\centering
\includegraphics[width=\textwidth]{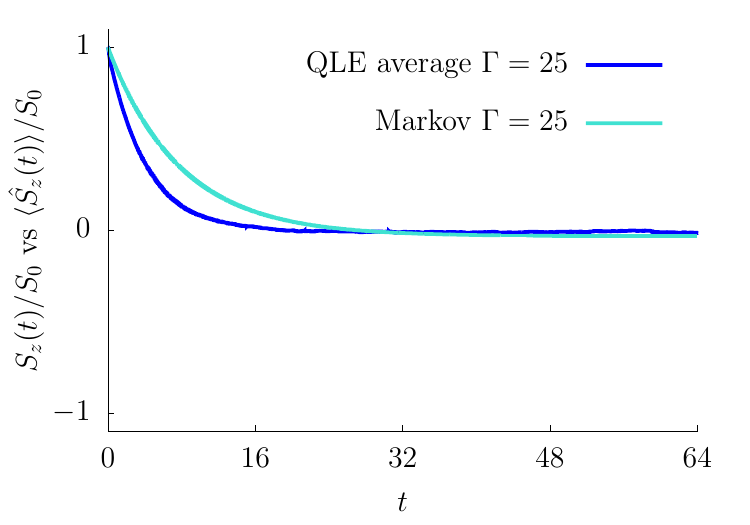}
\caption{The parameters chosen for the classical HL simulation are $\Gamma=25$, $\omega_0=12.5$ and $\alpha=2.5$ averaged over $25,000$ trajectories}
\label{fig:sub2}
\end{subfigure}
\hfill
\begin{subfigure}{5.5cm}
\centering
\includegraphics[width=\textwidth]{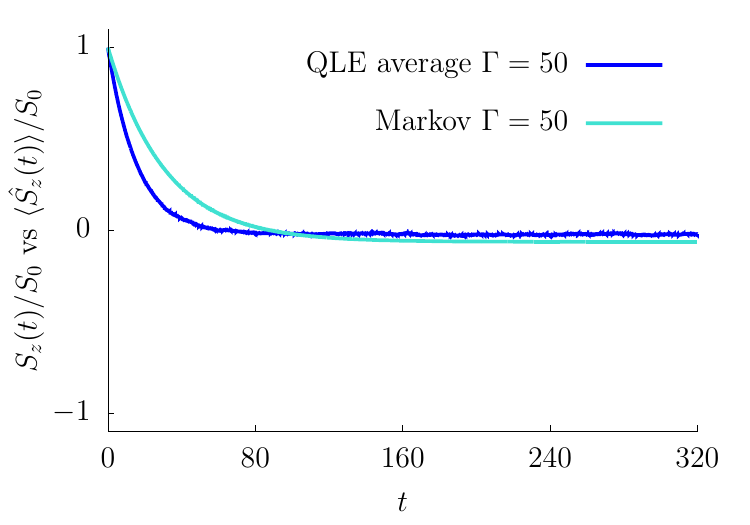}
\caption{In this plot, the HL was simulated with the parameters $\Gamma=50$, $\omega_0=25$, and $\alpha=5$. The curves are averaged over $25,000$ trajectories}
\label{fig:sub3}
\end{subfigure}
\caption{Here, the high temperature limit is plotted for three parameter choices at a temperature of $T=200$. The rapidly decaying turquoise curve originates from the classical HL simulation averaged over $25,000$ trajectories. The blue curve is the quantum mechanical expectation value computed along the lines of Eq. ~(\ref{HTEX}). The parameters chosen were $\Gamma_1=10$, $\Gamma_2=25$ and $\Gamma_3=50$, where all other parameters are dependently given by $\Gamma = 2\omega_0=10\alpha$. The parameters chosen yield dynamics moving deeper into the Markovian regime from left to right, since $\mu_{T,1}<\mu_{T,2}<\mu_{T,3}$.}
\label{fig:highT}
\end{figure*}

Having examined the comparison of quantum dynamics and classical HL simulation at $T=0$, we concluded that in this limit, neither the Markovian nor the non-Markovian configurations produce the correct steady state. One might therefore turn to the high-temperature limit to see if the issues present at $T=0$ are still present when the temperature is drastically increased. This, however, limits the scope of the investigation somewhat, since only the strictly Markovian quantum dynamics without any memory is firmly established for high temperatures.\\
The equation of motion governing the quantum mechanical time evolution of the probability amplitude for an excited $z$-component of a spin at high temperatures is laid out in the textbook by Scully and Zubairy \cite{scully1997quantum}. A firm derivation is given in the book by Breuer \cite{breuer2002theory}, established by means of projection operator techniques. The differential equation for the probability of finding such a system in the excited state $\psi(t)$ is similar to a modified version of Eq. (\ref{kernel1}). Here, strictly Markovian time evolution and an occupied bath are considered. This results in
\begin{equation}
    \dot{\psi} =  \lambda[\bar{n}-(2\bar{n}+1)\psi],
\end{equation}
involving the thermal average of the bosonic occupation number for the bath
\begin{equation}
    \bar{n} = \frac{1}{\exp \left(\frac{\omega}{T} \right) - 1}.
\end{equation}
Solving this differential equation and rescaling, one can see that the decay of the spin will follow
\begin{equation}\label{HTEX}
    \langle \hat{S}_z (t) \rangle = \frac{2(\bar{n}+1)}{2\bar{n}+1}e^{-(2\bar{n}+1)\lambda t} - \frac{1}{2\bar{n}+1}.
\end{equation}
This yields a more rapid decay and a convergence to a steady state differing from $-1$ that can be calculated according to 
\begin{equation}
    \lim_{t \rightarrow \infty} \langle \hat{S}_z(t) \rangle = -\frac{1}{2\bar{n}+1}.
\end{equation}
Having thus defined our reference for the high temperature limit, we can compare this decay to the classical HL average for a high temperature of $T=200$. The choice of parameters must be adapted to ensure that the classical HL simulation is deep in the Markovian regime, since the quantum calculation only holds for the strictly Markovian case. The condition for Markovianity must be adapted, since the bath now contributes strongly to the dynamics through high-temperature noise. At a nonzero temperature, the noise power spectrum from eq. (\ref{Power}) is enhanced by the factor $\coth(\omega/2T)$, which increases the spectral weight of fluctuations at frequencies that are resonant with the system. Since the decay rate is controlled by the noise power evaluated at the system frequency, this enhancement shortens the characteristic decay time scale without altering the spectral width that determines the bath memory time. In the high-temperature limit, $\coth(\omega_0/2T) \sim T/\omega_0$, therefore thermal noise accelerates decay by increasing the amplitude of the noise at resonance rather than by introducing a new dynamical time scale. Thus, the characteristic decay time scale for high temperatures is given by $\tau_\psi=\frac{\alpha}{4 \Gamma \omega_0} \coth \left(\frac{\omega_0}{2T} \right)$. This means that the system will be Markovian when the high-temperature Markovianity parameter $\mu_T$ fulfills
\begin{equation}
    \mu_T=\frac{2\Gamma^2\omega_0}{\alpha \coth \left(\frac{\omega_0}{2T} \right)} \gg 1. \label{HTMcond}
\end{equation} 
 In Fig. \ref{fig:highT}, we produced a comparison in the high temperature limit for three different parameter choices, corresponding to $\Gamma_1=10$ for \ref{fig:sub1}, $\Gamma_2=25$ for \ref{fig:sub2}, and $\Gamma_3=50$ for \ref{fig:sub3}. The remaining parameters are chosen according to $\Gamma=2\omega_0=10\alpha$ for all three cases, ensuring that $\omega_0$ is large enough, so that the RWA will hold. Here, the Markovianity parameters for the system decay in Eq. (\ref{HTMcond}) are $\mu_{T,1} \approx 12.5$, $\mu_{T,2} \approx 195.3$, and $\mu_{T,3} \approx 1560.5$ for the three different parameter sets. This implies that for each parameter set, we move about an order of magnitude deeper into the Markovian regime. Since the dynamics is dominated by noise at high temperature, smoothing out the curves requires heavier averaging; therefore, the number of realizations averaged over was raised to $25,000$. As can be seen by increasing $\Gamma$, the decay rates decrease as the parameter increases.
In this case, the presented approximated model at high temperatures slightly overestimates the decay rate compared to the exact quantum mechanical calculation. At high temperatures, the discrepancies with respect to the steady state become much less pronounced, compared to the limit $T=0$, yielding a similar long-time limit.

\section{Summary and Conclusion}\label{Sec6}

In conclusion, our goal was to investigate the validity of classical simulation of spin systems. In the work produced by Anders et al. \cite{anders2022quantum}, this approach is freely used, without a detailed justification, for the analysis of HL equations. This makes it inherently uncontrolled. Since the evolution of large open spin systems is computationally difficult, we benchmarked the results against analytically known quantum dynamics of a single spin-1/2 coupled to a bosonic bath.\\ 
To summarize, the comparison at zero temperature yields almost identical decay rates for both the Markovian set-up, as well as the non-Markovian set-up initially, compared to the average over many trajectories for the classical HL. However, the classically computed HL saturated quickly and plateaus at a steady state drastically differing from the true ground state, as shown in Figs. \ref{fig:average_decay} and \ref{fig:average_decay_2}. From WW theory, it is known that an excitation of a spin-1/2 at $T=0$ relaxes completely to the ground state when coupled to a bosonic bath. However, the classically computed HL has a thermal average, whose steady state is far away from the ground state for both the Markovian and non-Markovian configurations, due to persistent fluctuations around the ground state.\\
In the high-temperature limit, the only well-founded analytical quantum calculation considers the strictly Markovian case. This result was used as a reference case to benchmark the high-temperature dynamics of the classical HL simulation. Thereby, the discrepancies regarding the ground state were far less pronounced, see Fig. \ref{fig:highT}. For both the quantum mechanical result and the classical HL simulation, the decays are much faster and still decay on nearly identical time scales. However, the classical simulation features a decay rate a little bit larger than the analytical theory predicts.\\
For spins with a small spin quantum number in the low-temperature limit, this neglect of quantum features of the system in question crucially misrepresents the true steady state. In practice, this may not be a significant limitation to the applicability of this framework, as experimental setups focusing on the Landau-Lifshitz-Gilbert equation generally involve larger spins. Thereby, the classical ansatz may be better suited for spins with a larger quantum number. For such spins, quantum coherence and zero-point thermal fluctuations play a less crucial role. Therefore, the mismatch of the ground state will be less pronounced.\\ 
To that end, this study underscores the importance of benchmarking approaches that can reduce the richness of features of the system dynamics being considered in order to see what features are preserved and what features are lost.

\section*{Data Availability}

All our raw data have been deposited in the Zenodo database at https://doi.org/10.5281/zenodo.19115746 \cite{scott_2026_19115745}.

\section*{Acknowledgements}

We thank Janet Anders, Felix Hartmann, Jiaozi Wang, Robin Steinigeweg, and Stefan Linz for fruitful discussions. This work has been funded by the Deutsche Forschungsgemeinschaft (DFG), under Grant No. 531128043, as well as under Grant No.\ 397107022, No.\ 397067869, and No.\ 397082825 within the DFG Research Unit FOR 2692, under Grant No.\ 355031190.

\section*{Appendix: HL without any noise}

The power spectrum in Eq. (\ref{Power}) is constructed such that it follows a linear dependence with T at high temperatures but mimics quantum effects for low temperatures. One may consider a scenario with a purely classical power spectrum with a pure linear dependence in $T$, thus ignoring quantum zero-point fluctuations:
\begin{equation}
    \tilde{P}(\omega) = \pi \frac{T}{\omega^2}
\end{equation}
This would resolve the issue of the thermal average never reaching the ground state, since noise does not perturb the system around the ground state.
However, starting at $S_z(0)=+1$, we immediately find that this initial state is an unstable fixed point. By starting out in a fixed point, the spin remains at $S_z(t) = +1$ indefinitely. The intuition is that because the noise-free HL equation conserves spin length and the torque is orthogonal to $\hat{\mathbf{S}}$, initializing at the pole results in zero torque and hence no dynamics in $S_z$. We therefore see that a classical power spectrum directly causes issues when compared to the WW setup, when no perturbation of the initial condition is present. This issue of frozen dynamics can only be remedied if quantum zero-point noise is present at $T\rightarrow 0$ to dislodge the spin from the fixed point. It should furthermore be noted that the package \texttt{SpiDy} includes a noise generation function, where the zero-point fluctuations are removed by modifying the power spectrum according to
\begin{equation}
    \tilde{P}(\omega) = \begin{cases}
0, & T = 0, \\
\coth\!\left(\dfrac{\omega}{2T}\right) - \operatorname{sign}(\omega), & T \neq 0,
\end{cases}
\end{equation}
where $\operatorname{sign}$ is the sign function. As is evidently visible, this modification will also lead to frozen dynamics at $T=0$.

\bibliography{sample.bib}
\bibliographystyle{apsrev4-1_titles.bst}
\end{document}